\documentclass[11pt]{article}  
\usepackage{menuproc}
\usepackage{wrapfig}
\usepackage{cite}
\usepackage{epsfig}
\usepackage{amsmath,amssymb}
\begin{document}
%
%
%
\titlematter{Status Report on the Light Baryonic States}%
{B.M.K. Nefkens and J.W. Price}%
{UCLA, Los Angeles, CA 90095, U.S.A.}%
{The regularities in the spectrum of the light baryon resonances are
reviewed and compared with those of the light mesons.  We discuss the
occurrence of parity doublets and clusters, and note the trends in the
values of the masses, widths, spins, and parities.  The importance of
SU(3) flavor is illustrated and the status of quark model calculations
of the baryonic spectrum is reviewed.  The absence of evidence for
baryonic hybrids is particularly interesting.  We propose to use
better symbols for the baryon resonances which do not conflict with
the simple quark structure of QCD.  We shall comment also on fine
tuning the Star System for the hadronic states.  The importance of
greater support for the construction and operation of secondary beams
of $\pi$, $K$, $\bar p$, $\vec{n}$ and $\vec\gamma$ up to 5 GeV/$c$
for the future of non-perturbative QCD is emphasized.}%
%
%

\section{Introduction}
An important purpose of the biannual Symposium on ``Meson-Nucleon
Physics and the Structure of the Nucleon'' (MENU) is to review the
status of the light baryonic states.  MENU provides a public forum for
discussing the occurrence of regularities in the hadrons, and for
evaluating the success of various hadron models, particularly of the
light baryons which are made up of $u$, $d$, and $s$ quarks. 

\section{Patterns in the Widths of Baryons}
\begin{figure}[h!]
\centerline{\epsfig{file= 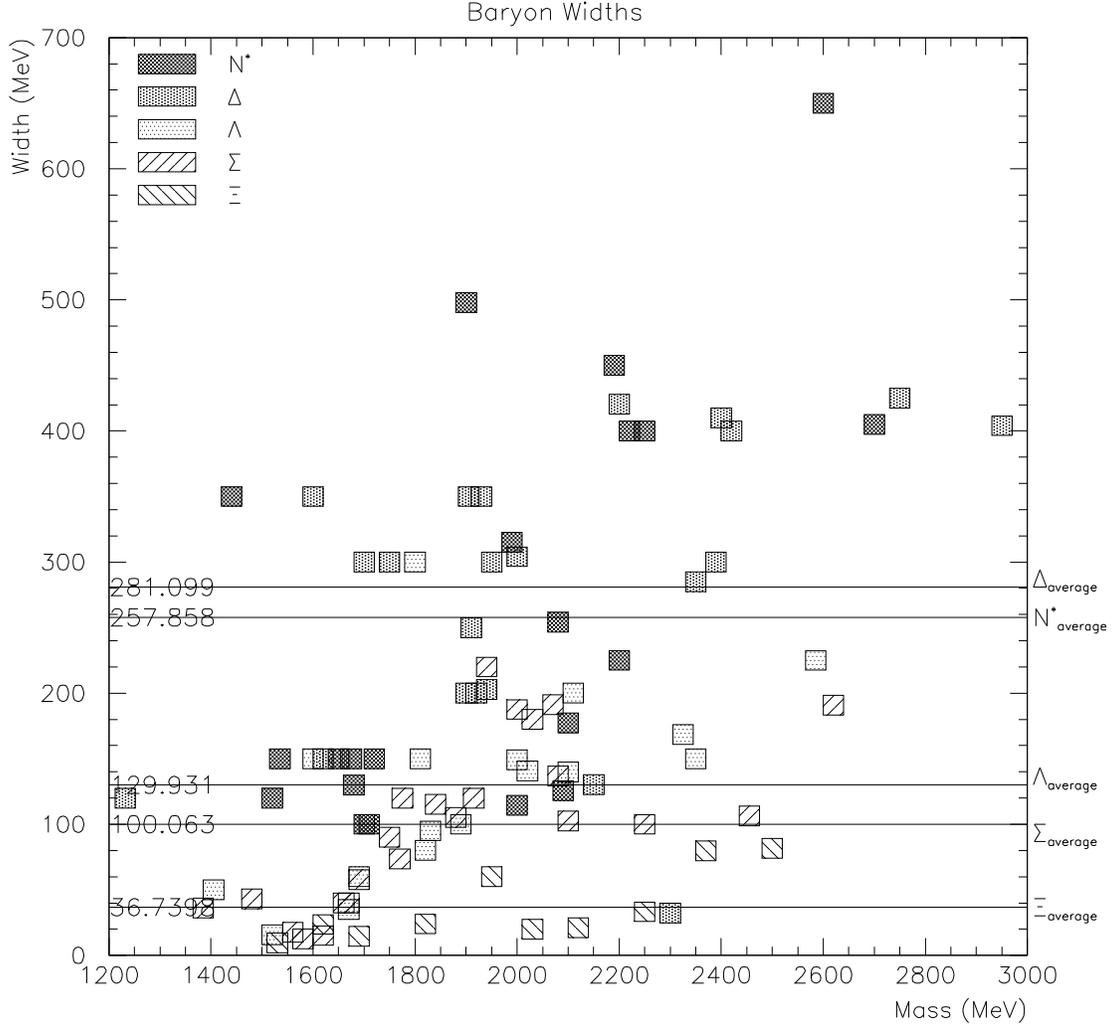,width=.9\textwidth,silent=,clip=}}
\caption{\label{fig:width}The width of all known light baryons.  The
horizontal lines are the average value for each family.} 
\end{figure}

The width, $\Gamma$, of all light baryon resonances as listed in the
Review of Particle Physics~\cite{PDG00} is shown in
Fig.~\ref{fig:width}. 

$\Gamma$ increases with the mass of the resonance and the magnitude
depends on the strangeness number of the family (which is directly
related to the number of $u$ and $d$ quarks.) The value of the average
$\Gamma$ for each family is given on the abscissa of
Fig.~\ref{fig:width}.  The relation between the widths is the
following: 
\begin{equation}
\Gamma(N^*) = 
\Gamma(\Delta^*) \simeq
\frac{9}{4}\Gamma(\Lambda^*) =
\frac{9}{4}\Gamma(\Sigma^*) \simeq
9\Gamma(\Xi^*).
\end{equation}
Riska~\cite{Riska} has noted that these ratios correspond to
$[\#(u+d)]^2$, where $\#(u+d)$ is the number of up and down quarks.
Note that the $\Xi ^*$ states are sufficiently narrow that they may be
fruitfully explored in production experiments such as $\gamma
p\to\Xi^-K^+K^+$; this provides a practical way for discovering many
of the missing $\Xi^*$ resonances. 

\section{Patterns in Baryon Masses, Spins and Parities}
\begin{figure}[h!]
\centerline{\epsfig{file=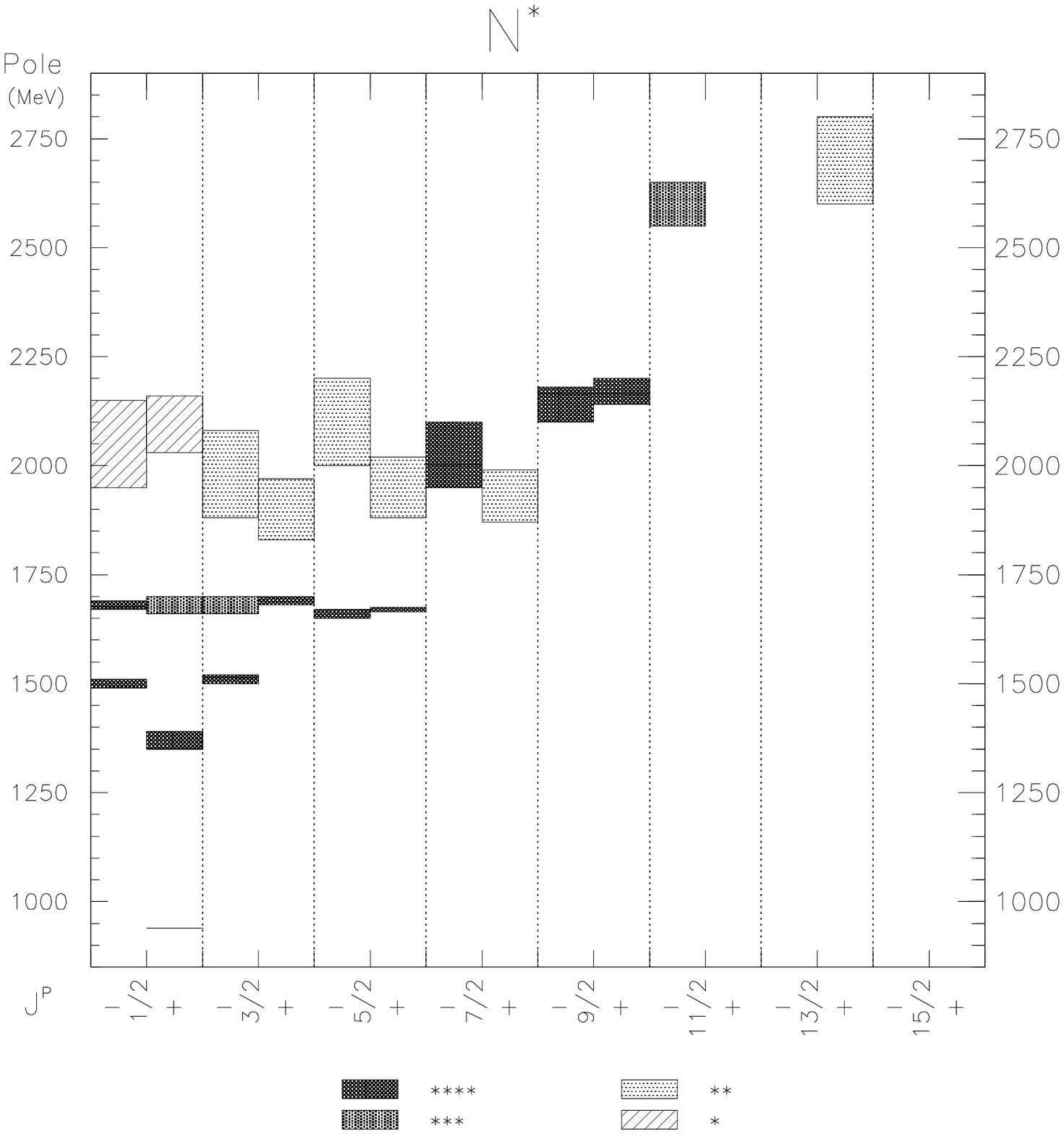,width=.7\textwidth,silent=,clip=}}
\caption{\label{fig:parity-pair}Parity pairing of the $N^*$ states.
Each resonance is plotted by its pole value in 8 columns for spin
$\frac{1}{2}$ to $\frac{13}{2}$ negative parity states are on the left
side and positive on the right.  The darkest shade of gray is for 4
star states.  The lightest is for one star.} 
\end{figure}

Shown in Fig.~\ref{fig:parity-pair} is a parity-pairing plot, which
displays by rectangular boxes the real part of the pole of every known
$N^*$ state in eight vertical bands, one for each spin: 1/2, 3/2,
$\cdots$ 15/2. 
Every band has two columns; the left for the negative parity states
and the right for the positive ones.  The star ranking of each state
is indicated by the shading: four stars (darkest shade) are awarded to
well-established states and one star (lightest shade) to the iffy
ones. 

There are clearly 3 mass regions in Fig.\ 2:  4 states have $m<1600$
MeV, where $m$ is the pole value of the state.  None of these has a
parity partner; we shall call them \emph{bachelor} states. There are
16 states with 1600 $< m <$ 2200 MeV which form 8 parity doublets, and
make two clusters.  The 2 remaining states at $m >$ 2500 MeV are
single states; however, the searches for other states have been far
from exhaustive.  A similar pattern of parity doublet states is found
in the $\Delta$ family.  For the case of the $\Lambda$ and $\Sigma$
states there are not sufficient data for drawing a firm conclusion
about the similar occurance  of parity doublets.

\begin{figure}[h!]
\centerline{\epsfig{file=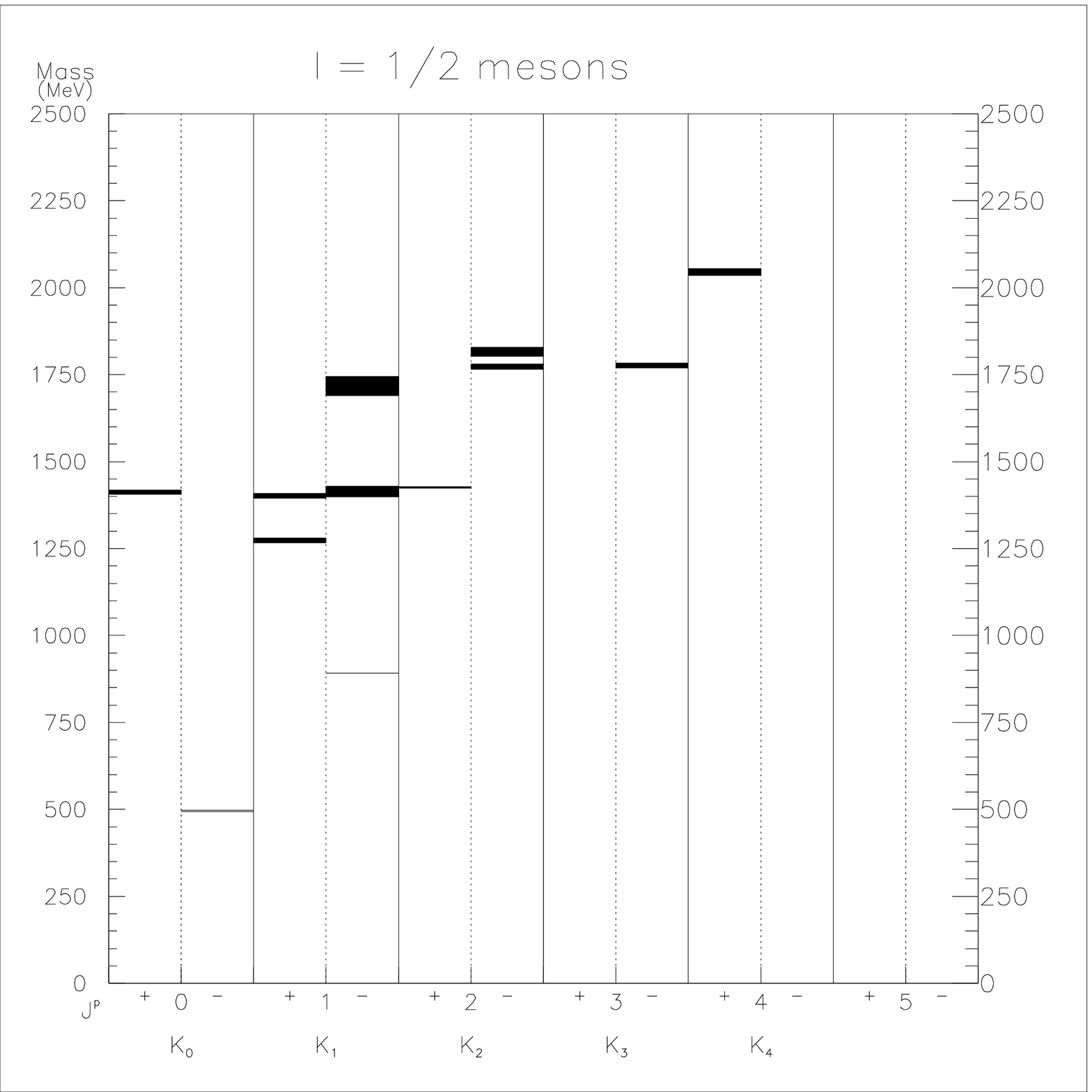,width=.7\textwidth,silent=,clip=}}
\caption{\label{fig:mesons}Lack of parity pairing of the light strange
mesons.}
\end{figure}

The situation for the mesons is fundamentally different.
Fig.~\ref{fig:mesons} shows the parity pairing plot for the strange
meson family. 
There is no evidence for parity doubling. The isosinglet and
isotriplet meson families support this.

We conclude that parity doubling is a feature of the baryons which is
\emph{not seen} in the mesons. Possible reasons for this could be a
diquark substructure~\cite{Iachello} or some hitherto overlooked
symmetry in the wave function.

\section{The Flavor Symmetry of QCD}
The Lagrangian of QCD, ${\cal L}_{QCD}$, is given by the following
compact expression~\cite{PDG00}:
\begin{equation}
{\cal L}_{QCD}=-\frac{1}{4}F_{\mu\nu}^{(a)}F^{(a)\mu\nu} +
               i\sum_q\bar{\psi}_q^i\gamma^{\mu}(D_\mu)_{ij}\psi_q^j -
               \sum_q m_{q}\bar\psi_q^i\psi_{qi}
\label{eq:LQCD}
\end{equation}
with
\begin{eqnarray*}
F_{\mu\nu}^{(a)} & = & \partial_\mu A^a_\nu - 
                       \partial_\nu A^a_\mu +
                       g_sf_{abc}A^b_\mu A^c_\nu,\\
(D_\mu)_{ij}     & = & \delta_{ij}\partial_\mu - 
                       ig_s\sum_a\frac{\lambda_{i,j}^a}{2}A_\mu^a.
\end{eqnarray*}
$\psi$ is the quark field, $A$ is the gluon field and $m_{q}$ is the
mass of quark $q$.  Eq.~\ref{eq:LQCD} may be arranged as follows:
\begin{equation}
{\cal L}_{QCD}={\cal L}_0 + {\cal L}_m.
\end{equation}
${\cal L}_0$ consists of the first two terms of Eq.~\ref{eq:LQCD} and
${\cal L}_m$ is the third term.  ${\cal L}_0$ depends only on the
fields; it is the same for all 6 quarks and 8 gluons.  This is the
famous flavor symmetry of QCD, which is a manifestation of the
universality of the strong interaction; it is broken by the mass term,
\[
{\cal L}_m = -\sum_q\ {\bar{\psi}_q}m_q\psi_q.
\]
The success of the SU(3) symmetry for systems of $u$, $d$ and $s$
quarks is indicative of ${\cal L}_0 >> {\cal L}_m$.  ${\cal L}_m$
produces a change of about 15\% in the mass of the hyperons.

Flavor symmetry explains the stunning similarity between the features
of threshold $\pi^-\ p\to\eta\ n$ production and $K^-p\to\eta n$ as
well as the amazing analogy between the Dalitz plots of $\pi^-\
p\to\pi^0\pi^0 n$ and $K^- p\to\pi^0\pi^0\Lambda$ and the
dissimilarity with $K^- p\to\pi^0\pi^0\ \Sigma^0$~\cite{Nefkens}.

\section{Baryon Mass Calculations}

\begin{figure}[h!]
\centerline{\epsfig{file=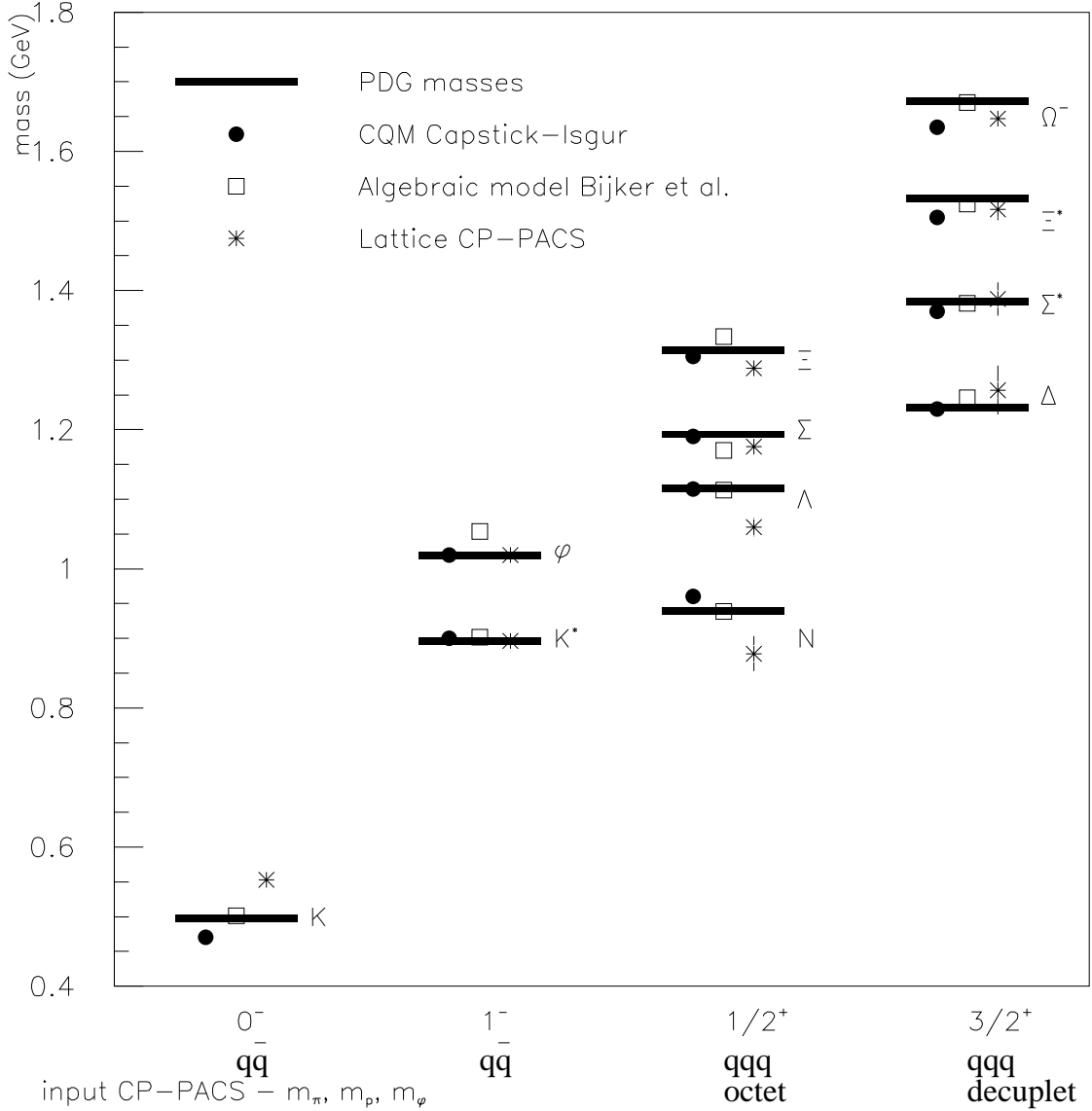,width=.9\textwidth,silent=,clip=}}
\caption{\label{fig:masses}Masses of light baryons.  The first column
shows the $K$-meson, the second column the $K^*$ and $\phi$. the third
column the baryon octet ground states, and the fourth column the
baryon decuplet ground states.  Solid lines are experimental data;
black dots are for Capstick-Isgur~\cite{Capstick}; open squares Bijker
et al.~\cite{Bijker}; stars CP-PACS~\cite{Aoki}.} 
\end{figure}

The experimental masses~\cite{PDG00} of the ground states of the four
light baryon-octet families, the $N$, $\Lambda, \Sigma$ and $\Xi$, are
displayed in Fig.~\ref{fig:masses} by thick horizontal lines. 
Shown also are the masses of the ground states of the four decuplet
families, the $\Delta$(1232), $\Sigma$(1385), $\Xi$(1530) and
$\Omega^-$(1672).  We shall compare these mass spectra with three very
different model calculations which are representative of the large
variety of calculations in this field. 
\begin{enumerate}
\item Lattice-gauge, L-G, results obtained by the CP-PACS
group~\cite{Aoki} for a quenched QCD calculations are indicated in
Fig.~\ref{fig:masses} by the stars.  The L-G calculations used as
input the masses of the $\pi^0, \rho$ and $\phi$ mesons, they also set
$m_u=m_d$.  The agreement of this L-G calculation with the
experimental masses is at the several percent level.  For example,
$m_p$(L-G) = 878 $\pm$ 25 MeV, while $m_p$(exp) = 938 MeV. One should
not be carried away by the level of agreement for the 8 baryon ground
states of this and other calculations; this could give an undeserved
sense of accomplishment.  Note that the masses of the four decuplet
states (which have a symmetric flavor state function) simply differ by
the $s-d$ quark mass difference.  Well before the birth of QCD, the
resulting equal mass spacings were known as the Gell-Mann decuplet
mass splitting  relation.  L-G and all quark models display this
decuplet relation to the level of 1 MeV, however, experimentally it
only holds to the level of 17 MeV.  A similar relation applies to the
octet ground-state masses, they obey the Gell-Mann-Okubo mass
relation.  This is different from the decuplet relation because the
octets have mixed flavor symmetry.  Thus, instead of 8 separate mass
values there are actually only 4 independent numbers: the mass of the
proton and the $p-\Delta$, $\Lambda-\Sigma$, and the $p-\Lambda$ mass
differences.  
\item The results of the relativized quark model calculations by
Capstick and Isgur~\cite{Capstick} are shown in Fig.~\ref{fig:masses}
by black dots.  The agreement with experiment is slightly better than
the L-G model. 
\item The algebraic model calculations by Bijker et al.~\cite{Bijker},
are shown by the open squares in Fig.~\ref{fig:masses}.  Again, the
agreement with experiment is excellent which in part originates in the
use of a larger input data set. 
\end{enumerate}

The above calculations are less satisfactory when it comes to
obtaining the mases of the baryonic excited states.  The limitations
inherent in using a quenched QCD calculations of the present L-G
models makes them not useful in their current form for the calculation
of the excited states.   We hope that this will change in the not too
distant future.

\begin{figure}[h!]
\centerline{\epsfig{file=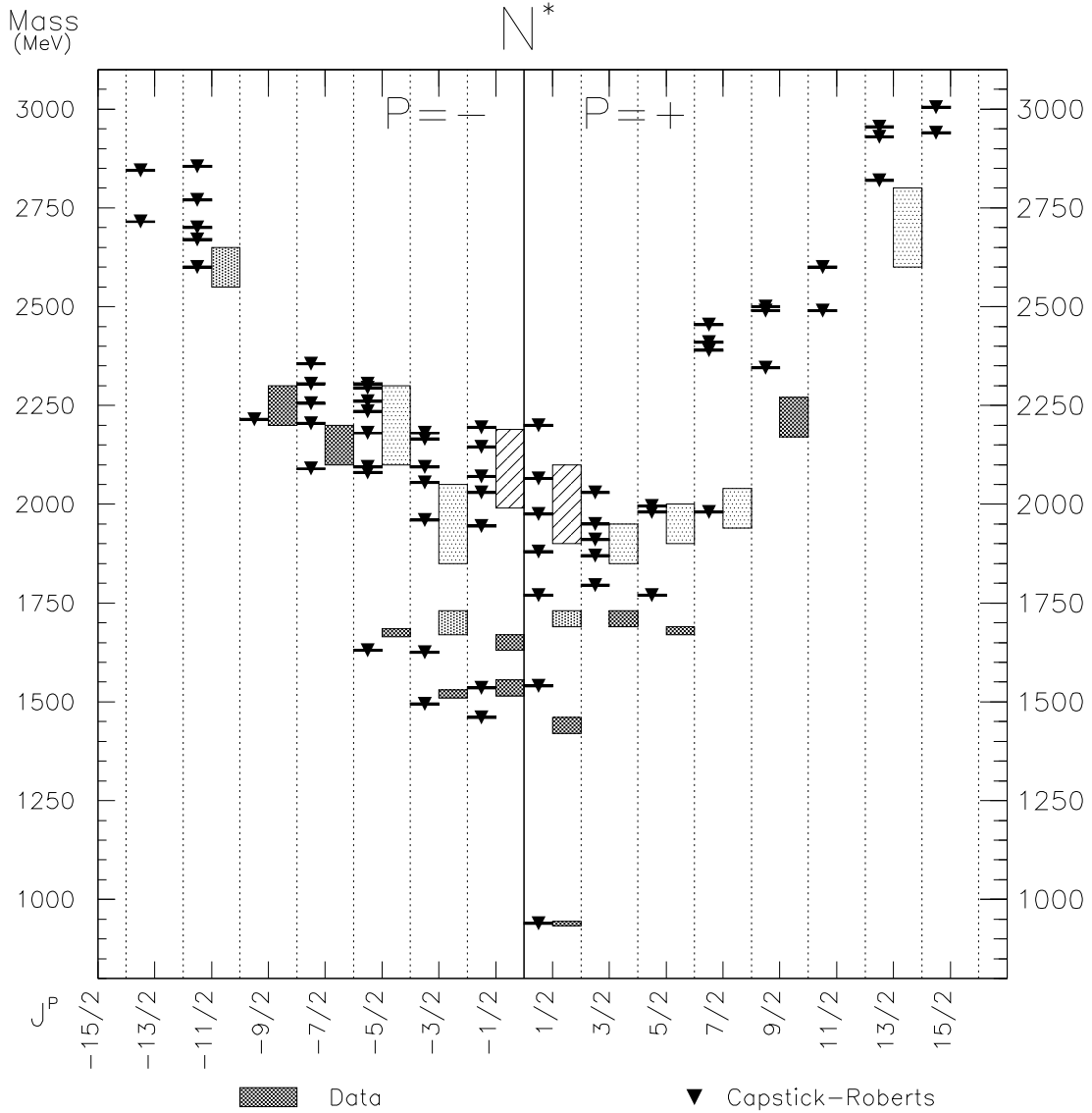,width=.9\textwidth,silent=,clip=}}
\caption{\label{fig:nstar-th-ex}$N^*$-mass spectrum.  The left side is
for negative parity states.  The right for positive.  The experimental
masses are given by the boxes.  The lines with the triangle in the 
middle are the calculation by Capstick and Roberts~\cite{Cap-Roberts}.} 
\end{figure}

Among the available quark model calculations we chose the
Capstick-Roberts~\cite{Cap-Roberts} work which comes from the same school as~\cite{Capstick}.  Shown in
Fig.~\ref{fig:nstar-th-ex} are the masses of all listed~\cite{PDG00}
$N^*$ states.  
We use different shades of gray to indicate the number of stars given
to each resonance.  The predictions by Capstick and Roberts are shown
by horizontal lines with a triangle in the middle.  Qualitatively, the
spectrum of the experimental masses below 2000 MeV is reproduced by
most quark models.  A close inspection reveals several nagging
discrepancies: 
\begin{enumerate}
\item The lowest established excited $N^*$ state is the Roper
resonance, which has positive parity and spin $\frac{1}{2}$, like the
nucleon which is the ground state.  In the quark model
of~\cite{Cap-Roberts} the lowest two states have negative parity and
spin $\frac{1}{2}$ and $\frac{3}{2}$.  This ordering is difficult to
rectify except by a major modification such as the direct
participation of Goldstone bosons in the quark-quark
interaction~\cite{Glozmann}. 

\item The calculated masses of the positive parity states are all too
high by some 80 MeV compared to the data, while all negative parity
resonances are calculated to be too low by some 40 MeV.  It is
interesting that a similar quark model calculation~\cite{Godfrey} of
the mesons agrees very well with the data. 

\item Less than a quarter of the predicted states above 2000 MeV have
been observed experimentally.  The reason which is usually advanced
for not seeing the ``missing resonances" is their small coupling to
the $\pi N$ channel used for the identification by the $\pi N$ partial
wave analyses.  A small coupling is indeed a feature of several quark
models which use an independent channel calculation of the $\pi N$
branching ratio.  In reality there is a non-negligible coupling
between various channels such as the $\pi N$ and $\eta N$ channel.  In
these cases the $\pi N$ final state is enhanced because it has the
larger phase space. 
\end{enumerate}
Experimentally it will be hard to identify the plethora of missing
$N^*$ states with a mass $>$ 2000 MeV.   According to
Fig.~\ref{fig:width} we expect these states to have a width $>$ 300
MeV.  In the region 2000--2300 MeV the quark model predicts 30 states,
all overlapping and broad. We propose that the mystery of the missing
baryonic resonances be settled by a detailed investigation of the
excited $\Xi$ because all $N^*$ states are related by flavor symmetry,
discussed in section 4, to $\Xi$ states that have the same spin and
parity and a $\sim$450 MeV larger mass.  However, they have a narrow
width of $\sim$ 40 MeV.  The $\Xi^*$ states are readily accesible in
production experiments such as $K^-p\to K^+\Xi^*$ and $\gamma p\to
K^+K^+\Xi^*$. 

\section{Where are the Hybrid Baryons?}

\begin{figure}[h!]
\centerline{\epsfig{file=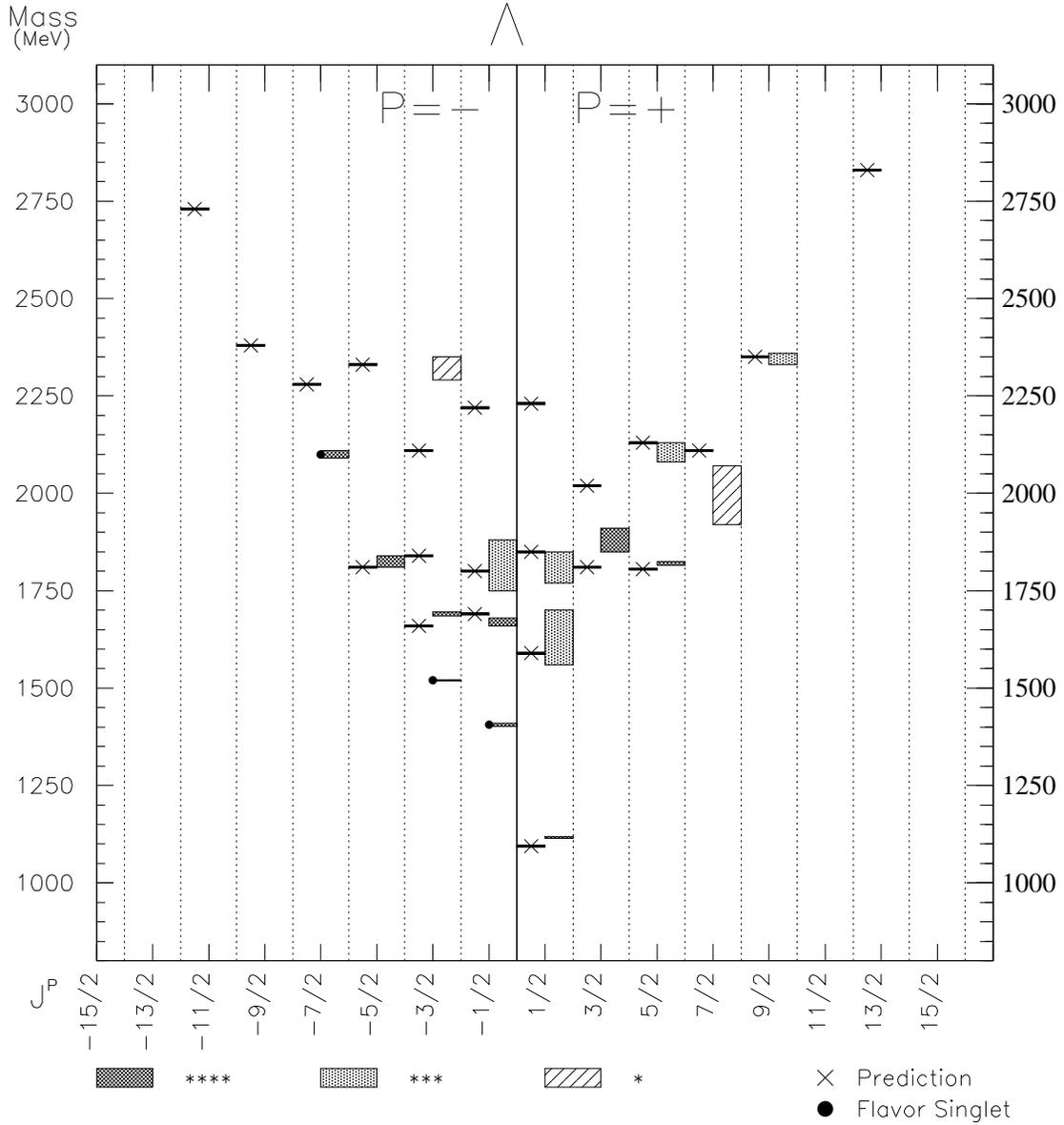,width=.9\textwidth,silent=,clip=}}
\caption{\label{fig:lambda_th_ex}$\Lambda^*$-mass spectrum.  The left
side is for negative parity states, the right for positive.  The
experimental $\Lambda^*$ masses are given by the boxes.  The crosses
are the theoretical predictions based on the known $N^*$mass
values~\cite{Nefkens} and flavor symmetry.} 
\end{figure}

There is no argument known which is based on QCD or on our
understanding of confinement for limiting  the baryons to 3 quark
states, $|B\rangle=|qqq\rangle$.  We expect also
$|B\rangle=|qqqg\rangle$, $|B\rangle=|qqqgg\rangle$, etc.  The latter
two are called the hybrid baryons.  They do not follow the simple
SU(3)-flavor symmetry relations between the different light-quark
baryon families.  Thus, we do not expect that an $N^*$ hybrid will
have a flavor partner in the $\Lambda$ family and vice versa.  Shown
in Fig.~\ref{fig:lambda_th_ex} by boxes are the various known
$\Lambda$ states using gray shading to indicate their star rating.  
We also show by the horizontal lines with crosses the SU(3) flavor
prediction based on the experimentally observed $N^*$ states and a
simple expression to adjust for the flavor breaking due to the $s-d$
quark mass difference~\cite{Nefkens}.  From this figure we can
conclude that:
\begin{enumerate}
\item The Roper resonance, the $N(1440)\frac{1}{2}^+,$ which has long
been regarded as a hybrid candidate is not a hybrid because of the
existence of the SU(3) flavor partner the $\Lambda(1600)\frac{1}{2}^+$.

\item There are no unaccounted-for $\Lambda^*$ states, hence, there
are no $\Lambda$ hybrid candidates.
\end{enumerate}

\section{Nomenclature}

To identify a specific baryonic resonance many authors use an outdated
nomenclature which can be rather misleading.  In this outdated system
the proton is labelled a P$_{11}$ state implying one unit of angular
momentum of the constituents.  This label dates back to the meson
period in the history of nuclear physics when the nucleon was
considered to have a Dirac core surrounded by a P-wave pion cloud.
The latter ingredient was needed to account for the observed magnetic
dipole moments of the nucleons, $\mu_{p}\ =\ +2.79\mu_{B}$ and $\mu_n\
=\ -1.93\mu_B.$ 

In the quark era, with the success of QCD, it is hard to support the
notion that the proton, which is the ground state of the $N^*$ family
and has a life time (into certain channels) in excess of 10$^{32}$
years, is a P-state.  We can rectify the situation by dropping the
misleading nomenclature of $L_{2I2J}$ and replace it by a simple
system that uses measured parameters only.  The first part is a
capital or greek letter for the unique identification of the six light
baryon families, the $N,\Delta,\Lambda,\Sigma,\Xi,$ and $\Omega$.
This is followed by the mass in brackets and by the spin and parity.
Thus, the proton's new symbol is $N(938)\frac{1}{2}^+,$ the Roper is
$N(1440)\frac{1}{2}^+,$ the D$_{13}$ is $N(1520)\frac{3}{2}^-,$ and
the S$_{11}$ is $N(1535)\frac{1}{2}^-$. Similar conventions are used
for the other families, e.g.\ the $\Lambda$ ground state is
$\Lambda(1116)\frac{1}{2}^+,$  the lambda-Roper is
$\Lambda(1600)\frac{1}{2}^+,$ etc. 

\section{The Star System}

A practical system for quality assessment based on awarding a number
of stars --- as done by a well known restaurant guide --- has been in
use in baryon spectroscopy for many years.  Every baryon resonance
listed in the Review of Particle Physics~\cite{PDG00} is awarded 1 to
4 stars.  The meaning of the number of stars is the following. 
\begin{itemize}
\item[****]Existence is certain, and properties are at least fairly
well established. 

\item[***]Existence ranges from very likely to certain, but further
confirmation is desirable and/or quantum numbers, branching fractions,
etc. are not well determined.

\item[**]Evidence of existence is only fair.

\item[*]Evidence of existence is poor.
\end{itemize}
This system works well for the $N$ and $\Delta$ families where all
states have been investigated in several full, energy dependent and
independent, $\pi N$ partial wave analyses (PWA).  The 3 and 4 star
$\Lambda$ and $\Sigma$ states are in good shape as they come mainly
from $\bar{K}N$ PWAs.  However, there are several unsavory 1 and 2
star $\Sigma$ candidates, which represent some questionable bumps in a
few production experiments into inelastic channels.  A major problem
occurs in the case of the heavy baryons where all states have been
discovered in production experiments.  None of the new heavy baryons
have an experimental determination of their spin and parity; instead,
they are assigned a value based on the predictions of some popular
quark models.  We have seen in Sect. 5 how even the most extensive and
widely used quark model does a poor job in the mass ordering of the
lowest excited states.  Furthermore, the actual mass calculation
especially of the positive parity states is inaccurate by up to 80
MeV. Yet, the heavy baryon states have been given 3 and a few even 4
stars.  The spin and parity are the vital characteristics of any
resonance and a state does not warrant 3 or 4 stars when their is no
experimental data on its spin and parity.  We should fine tune the
definition of 3 stars with this in mind. 

Mark Manley~\cite{Manley} is floating the idea that we should
establish a new class of 5 star states which is reserved for the
\lq\lq golden" resonances about whose existence and basic quantum
numbers and properties there is no question.  This idea has merits and
deserves careful consideration by our community.  In the meantime all
physicists should be aware that the spin and parity of all heavy
baryons are assigned based on the quark model without experimental
verification. 

\section{Summary and Conclusions}

A large body of detailed information on the properties of the light
baryons~\cite{PDG00} has been accumulated.  However, our knowledge is
still very incomplete; it is insufficient to allow drawing reliable
conclusions about the occurance of significant regularities such as
parity doublets and clusters.  Investigating the occurance of
regularities is needed to make progress on the problem of quark
confinement in QCD. There is currently no evidence for the existence
of hybrids but we cannot exclude them either. The lack of any clear
manifestations of the gluon degree of freedom in any baryonic system
is unsettling.  It points to hitherto unexpected aspects of QCD in the
non-perturbative regime.

New data on the properties of the many expected, but undiscovered
$\Lambda^*,\Sigma^*,\Xi^*$ and $\Omega^*$ states are urgently needed
so we can establish  the dependence of the $s-d$ quark mass difference
on the energy and spin/parity of the confined 3 quark system.  A
convenient way to handle this is by investigating the validity of the
Gell-Mann decuplet and the Gell-Mann-Okubo octet mass relations for
high mass states with large spin and for the positive as well as
negative parity states.  This is needed for progress in the area of
the \lq\lq Origin of Mass", one of the areas of importance in our
field.  A coordinated effort is needed on the existence of $N^*$ and
$\Delta^*$ resonances with $m>$ 2000 MeV. Required for this are
sophisticated detectors and secondary beams of $\pi^\pm, K^\pm,
\bar{p}, \vec{n},$ and $\vec{\gamma}$ up to 5 GeV.  It is the
responsibility of this   community to raise the awareness of our
colleagues to the importance of the physics we are engaged in and to
the experimental tools, especially the secondary beams, required to
get our jobs done.


\end{document}